\magnification=\magstep1
\def\buildurel#1\under#2{\mathrel{\mathop{\kern0pt #2}\limits_{#1}}}
\input amstex
\documentstyle{amsppt}
\loadbold
\openup6\jot
\input amstex.tex
\input amssym.tex
\def\buildurel#1\under#2{\mathrel{\mathop{\kern0pt #2}\limits_{#1}}}
\def\RR{R\!\!\!\!\!\!I\,\,\,}

\def\i{I\!\!\!I\,}
\def\e{\epsilon}
\def\t{\tau}
\def\E{E^{\rho}}

\def\e{\epsilon}
\def\d{\delta}

\def\l{\lambda}

\def\g{\gamma}

\def\r{\rho}
\vskip.01truein
\centerline{\bf Self Diffusion in Simple Models:}  
\centerline{\bf Systems with Long Range Jumps.}
\bigskip
\centerline{by}
\bigskip
\centerline{A. Asselah}
\centerline{Department of Mathematics}
\centerline{Rutgers University}
\centerline{New Brunswick, NJ 08903}
\medskip
\centerline{R. Brito}
\centerline{Departamento Fisica Aplicada I}
\centerline{Facultad de CC. Fisicas}
\centerline{Universidad Complutense}
\centerline{28040--Madrid, Spain}
\medskip
\centerline{and}
\medskip
\centerline{J.L. Lebowitz}
\centerline{Department of Mathematics and Physics}
\centerline{Rutgers University}
\centerline{New Brunswick, NJ 08903}
\bigskip
\bigskip
\noindent
{\bf Abstract}

We review some exact results for the motion of a tagged particle in simple
models. Then, we study the density dependence of the self diffusion
coefficient, $D_N(\rho)$, in lattice systems with simple symmetric
exclusion in which the particles can jump, with equal rates, to a set of
$N$ neighboring sites.  We obtain positive upper and lower bounds on
$F_N(\rho)=N((1-\r)-[D_N(\rho)/D_N(0)])/(\rho(1-\rho))$ for $\rho\in
[0,1]$.  Computer simulations for the square, triangular and one
dimensional lattice suggest that $F_N$ becomes effectively independent of
$N$ for $N\ge 20$.

\bigskip
\bigskip

{\bf 1.  Introduction}

Many properties of macroscopic systems are universal, 
retaining their qualitative features under drastic
simplifications of the underlying microscopic structures.
Thus, lattice gas models have greatly enhanced our 
understanding of phase transition phenomena in equilibrium systems.  
Their dynamical behavior, currently an active area of research,
promises to be similarly fruitful for understanding nonequilibrium
properties of macroscopic systems.  

This article explores some aspects of
self-diffusion in lattice models. After a brief overview
of some rigorous results, we derive new results for systems with
long range jumps.
It is dedicated to Matthieu H. Ernst, a leader in the field of kinetic 
theory and lattice gases, on the occasion of his sixtieth birthday.

We shall be concerned here with the motion of a tagged particle in an
infinite interacting particle system.  A tagged particle is {\it exactly}
like any other particle in the system, its `tag' permits us to follow its
trajectory $X(t)$. This yields a relatively simple probe of time
correlations in a system of interacting particles in an overall stationary
state.

The self-diffusion coefficient $D_s$ is defined, in an infinite stationary
system without drift, as [1]
$$
D_s={1 \over 2d} \lim_{t \to \infty} {1 \over t} 
\langle [X(t) - X(0)]^2\rangle,
\eqno(1.1)
$$
where $d$ is the spatial dimension of the system and the average $\langle
\rangle$ is over the stationary measure.  We expect that in a real fluid
the limit (1.1) will exist, be positive, and be given by the
Einstein-Green-Kubo formula
$$
D_s = {1 \over d} 
\lim_{t\to \infty}\int_0^t \langle {v}(\tau) \cdot {v}(0)
\rangle d\tau,
\eqno(1.2)
$$
where 
$\langle {v}(\tau) \cdot {v}(0) \rangle $
is the velocity autocorrelation function [1]:  A simple computation
gives $\langle (X(t) - X(0))^2 \rangle = 2 \int_0^t (t-\tau) \langle
{v}(\tau)\cdot {v}(0) \rangle d\tau$, so (1.2) reduces to (1.1)
when $\langle {v}(\tau) \cdot {v}(0) \rangle$ decays
sufficiently rapidly.

The self-diffusion coefficient is a 
global dynamical parameter associated with 
macroscopic system in equilibrium, i.e. spacially uniform. 
Therefore, it is generally different from the bulk
diffusion coefficient $D_b$, which relates to the
evolution of a nonuniform density in a non-stationary system.  
$D_s$ can be thought of as a 
color diffusion coefficient by considering the 
evolution of the relative density of
two components of a system which differ only by
a property, say color, that plays no role in the dynamics, while 
the overall system, ignoring color, is in a uniform state [1, 2]. 
An approximate experimental realization of such a situation occurs
when the components are isotopes of 
$He^3$ atoms whose spins are polarized in different directions.

Going beyond (1.1) and (1.2), we can
introduce a ``scaling'' parameter $\epsilon$ and define
$X_\epsilon(t)$ as $\epsilon[X(t/\epsilon^2;\cdot)-X(0)]$,  where 
the $\cdot$ indicates the dependence of the trajectory on the coordinates
and velocities of {\it all} the particles at $t=0$ [1,2].  
Typically, we expect that in the long time limit,  $\epsilon \to 0$, 
the process $\{X_{\e}(t),\ t\in \RR\}$ converges in probability
-after subtracting out any drift- to the law
of a Brownian motion $\{W_{D_s}(t),\ t\in \RR\}$ with diffusion
coefficient $D_s$ given by (1.1)[1,2]. We summarize this by
$$
\lim_{\e\to 0}X_\epsilon(t)= W_{D_s}(t).
\eqno(1.3)
$$

The behavior (1.1)-(1.3) has been proven for the one component, one
dimensional system of hard rods with diameter $a$ [1, 3, 4].  For this
idealized system $D_s$ can be computed exactly,
$$
D_s(\rho) = {(1 - \rho a) \over \rho} \langle |v| \rangle, 
\quad \text {and} \quad
\langle |v| \rangle = \int^\infty _{-\infty} |v| h(v)dv.\eqno(1.4)
$$
Here $h(v) = h(-v)$ is the one particle velocity distribution function;
this need not be Maxwellian, since collisions in this system merely
exchange velocities.  Noting that $(1/\rho-a)$ is the mean free path in
this system, the interpretation of (1.4) is very simple.  On the other
hand, the velocity autocorrelation function $\langle v(\tau) v(0)
\rangle$ depends non-trivially on $h(v)$:
it decays like an exponential when $h(v)$ vanishes
near $v=0$, and like $t^{-3}$ when  $h(v)$ is Maxwellian [5].

The only other continuum system for which 
the existence of the limits (1.1)-(1.3) has been proven 
is a system of interacting Brownian particles ([6,7])
which models suspensions
of polymers or even of small macroscopic balls in a fluid. 
Actually, one needs to assume ergodicity of the dynamics, and
formula (1.2) has to be modified because instantaneous 
velocities are no longer well defined [1].

In $d=1$, stochastic models in which
the particles cannot cross each other
behave differently from the mechanical model which yields
(1.4): $\langle X^2(t) \rangle \sim \sqrt t$ so $D_s = 0$ [1,8].
Interestingly, however, $\sqrt \epsilon X(t/\epsilon^2;\cdot)$ 
still goes to a Gaussian process; see [4] for a simple derivation of 
the one dimensional results.  

\noindent 
{\bf 2. Lattice Models}
\medskip
{\bf a)  General Dynamics}

We consider now systems with one type of particles whose
total number is the only quantity conserved by the dynamics.
We expect however that much of our discussion will remain valid also
for models where momentum is also conserved [7, 9].
The stochastic dynamics of these systems consists of particles
``jumping''  between lattice sites.  The jump  from a site ${x}$ to a
site ${y}$ on the lattice occurs with a rate $c({x}, {y};
{
{\eta}})$, where ${{\eta}}$ is the configuration of the system
just 
prior to the jump:  ${\eta}
= \{{\eta} ({z})\}$, with ${\eta}({z}) = 0,1,2,..$, 
specifying the number of particles at site ${z}$.  
We shall generally consider the $d$-dimensional (simple) cubic lattice
${\Bbb Z}^d$.  

The system will be in a stationary state with measure
$\nu$ whenever
$$
\sum_{x,y} c(x,y;{\eta}) \nu({\eta})
= \sum_{{x},{y}} c({x},{y}; {\eta}^{x,y}) \nu({\eta}^{{x},{y}}),
\eqno(2.1)
$$
where $\eta^{x,y}$ is the configuration which arises from
${\eta}$ after a particle has jumped from $x$ to $y$.
A simple way to satisfy (2.1) is to have the equality hold
for each term in the sum. The rates
are then said to satisfy detailed balance with respect to $\nu$.  
In such cases $\nu$ can 
be written in the form of a Gibbs measure, $\nu_{eq}({\eta})
\sim \exp[-\beta H({\eta})]$, where $H({\eta})$ is
the energy of a configuration ${\eta}$, and $\beta$ is the
reciprocal temperature:  see [10] for a detailed discussion of 
Gibbs measures.  Detailed balance then corresponds to 
$$
c(x,y;\eta)/c(y,x;\eta^{x,y})=\exp\{-\beta[H(\eta^{x,y})-
H(\eta)]\}.
\eqno(2.2)
$$
In the probability literature a stationary process whose rates satisfy
detailed balance is called reversible: a film of the system in the
stationary state would look the same if run backwards.  

The trajectory $X(t)$ now takes
values on the lattice.  However, after scaling with $\epsilon$ and letting
$\epsilon \to 0$, as in (1.3), the limit will again be a continuous process.

\medskip
{\bf b)  Models without Exclusion}

One of the simplest dynamics for a system of particles on a lattice
is the so called ``zero-range'' process [11].  This corresponds to the
jump rates $c({x}, {y}; {\eta})$ depending 
only on $\eta({x})$, the number of particles at site ${x}$,
$$
c({x}, {y}; {\eta}) = \lambda g(\eta({x}))
p({y} - {x}).\eqno(2.3)
$$
Here $\lambda$ is a constant, $\lambda > 0$, while $g$ and $p$ satisfy the
conditions
$$
g(0) = 0, \quad \quad g(k) > 0, \quad \quad for \quad k > 0,\eqno(2.4)
$$
$$
p({0}) = 0, \quad \quad p({r}) \geq 0, \quad \quad \sum_{{r} \in {\Bbb Z}^d}
p({r}) = 1.\eqno(2.5)
$$  

The stationary measures for this dynamics in the macroscopic (infinite
volume) limit, are a translation invariant family of product measures,
$\nu_{\rho}$, 
parametrized by the average density $\rho$[11].  
The probability of having exactly $j$ particles at any given site is 
$$
W_j = {b^j \over G(j)} W_0,\quad \quad j = 0,1,2,...\eqno(2.6)
$$
where
$$
G(0) = 1, \quad \quad G(j) = \Pi^j _{l=1} g(l), \quad \quad j \geq 1,\eqno(2.7)
$$
and the parameters $b$ and $W_0$ are determined by the normalization and
the specified average density, $\rho \geq 0$, i.e. 
$$
\sum^{\infty}_{j=0} W_j = 1,
\qquad \text {and}\qquad 
\sum^{\infty}_{j=0} j W_j = \rho.\eqno(2.8)
$$
An easy check shows that these measures satisfy the detailed balance
condition (2.2), with $\beta H({\eta})$ a sum of single site
energies equal to $-\log W_j$, if and only if $p(r) = p(-r)$.  

Two particular cases of the zero range process deserve mention.  When 
$g(l) = l$ the dynamics corresponds to that of independent particles.  
This gives rise to the Poisson distribution
$$
W_j = (\rho^j/j!) e^{-\rho},\eqno(2.9)
$$
Taking $g(l) = 1 - \delta_{0,l}$, corresponding to only the `top' particle
jumping,  yields a geometric stationary distribution
$$
W_j = {1 \over {1+\rho}}({\rho \over 1+\rho})^j.\eqno(2.10)
$$

The stationary measure seen from the tagged particle is the `Palm measure'
$$
\hat \nu_{\rho}(\eta)={\eta(0)\over \rho} \nu_{\rho}.
$$
As the ``waiting time'' of the tagged particle depends on the number of
particles at the same site, its average jump rate is given by 
$$
\bar \l= \sum^\infty_{k=1} {\l_k \over \rho}W_k,
\qquad\text {with}\qquad
\l_k =\lambda g(k).
\eqno(2.11)
$$

Let the displacement, after $K$ steps, of the random walk specified
by transition probability $p(r)$ be $X_K$.
Assuming for simplicity that there is no drift, 
$$
\sum {r} p ({r}) = 0,\eqno(2.12)
$$
we have
$$
\langle X^2_{K+1} \rangle = \langle (X_K + Y_K)^2
\rangle = \langle X^2_K \rangle + \langle Y^2_K\rangle,
\eqno(2.13)
$$
where $Y_k$ is the displacement of the particle at the $(K+1)$ step.  
Clearly,
$$
\langle Y_K^2 \rangle = \sum {r}^2 p ({r}) \equiv 
2{\tilde D}_0,\eqno(2.14)
$$
and
$$
\langle X^2_K \rangle =2{\tilde D}_0K.
\eqno(2.15)
$$
A little thought shows that for the zero range process, the trajectory of
the tag will look the same as the trajectory of a single particle
performing a random walk on the lattice with transition probabilities
$p({\bold r})$.  The only difference is that the ``waiting time'' at any
site will generally depend on the number of particles there. In fact [9],
As soon as the process is ergodic, the scaled trajectory $X_\epsilon (t)$
satisfies (1.3) with
$$
D(\rho) = \bar {\l}(\rho){\tilde D}_0.
$$
Ergodicity is easy to show for $g(k)=k$, and was shown 
in the case $g(k)=1-\delta_{k,0}$ in [12]. In fact,
(1.3) is proven in [13] for all $g(k)$.

For the case of independent particles,
$\bar \l$ is independent of $\rho$ and equal to $\l$, while for
$g(k)=1-\delta_{k,0}$,
$$
\bar {\l } (\rho) = {\lambda\over 1+\rho},
$$
so that $D(\rho)$ decreases with density for this case.  The opposite
behavior is clearly also possible.  

Looking back on our arguments leading to (2.15) we see that the main
ingredients are the independence of the step $Y_K$ from the past
history of the process (e.g. $X_k$ is a martingale). 
This means that (2.15) should remain valid whenever
$$
c({x}, {y};{\eta}) = h({\eta}; {
x}) p({y} - {x}),\eqno(2.16)
$$
i.e.\ as long as $c({x}, {x} + {r}; {
\eta})/c({x}, {x}+{r}^\prime; {\eta})$ is
independent of ${\eta}$ (and by translation invariance of
${x}$).  

A particular example of (2.16) is (a generalization of) a model due to van
Beijeren [14], in which
$$
c({x}, {y}; {\eta}) = \sum_{i=1}^k \lambda_i
g_i(\eta({x})) p_i({y} - {x}),
$$
with the $g_i$ and $p_i$ satisfying the conditions (2.4),
(2.5) and (2.12).  The stationary measure is now not known in general.
In fact we expect that it will have very long range
correlations [14], yet (1.1) and (1.3) should still be valid with 
$$
{\tilde D}(\rho)={1\over 2}\bar \l(\r) 
\sum_i r^2 p_i(r)
$$
where $\bar \l(\r)$ is the average rate in the stationary measure.

{\bf Remark}:  It is clear that the diffusion constants $D$ and
${\tilde D}_0$
are, for anisotropic $p({r})$, the traces of the corresponding
positive-definite diffusion tensors ${D}$ and ${D}_0$.
When (2.12) holds and $p({r})$ is isotropic with respect to the
lattice directions then $D$ is diagonal.
\medskip
{\bf c)  Models with Exclusion}

We consider now the case where there is a hard core interaction between the
particles, forbidding the presence of more than one particle at any lattice
site.  The configurations of the system are now given by ${
\eta} = \{\eta({x})\}$ with $\eta({x}) = 0$ or $1$ and
$c({x}, {y}; {\eta}) = 0$ when $\eta({y}) =
1$.  The simplest dynamics for these systems correspond to the so called
simple exclusion processes, in which the jump rate 
from a site ${x}$ to a site ${y}$ is independent of the
configuration at other sites of the lattice,
$$
c({x}, {y}; \eta) = \lambda \eta ({x})(1 - \eta({
y})) p({y} - {x}),
$$
with $p({r})$ satisfying (2.5).

The translation invariant stationary measures $\nu_\rho$, with $0 \leq \rho
\leq 1$, will again 
be product measure with
$$
W_0 = 1 - \rho, \quad \quad W_1 = \rho, \quad \quad W_j = 0 \quad for \quad
j \geq 2.
$$
The $\nu_\rho$ will, like before, satisfy detailed balance if and only if
the jump rates are symmetric, $p({r}) = p(-{r})$.

The behavior of a tag trajectory $X(t)$ is now considerably more
complicated than in the zero range process.  In particular,
knowledge of the past history of
$X(t)$ will influence the probabilities that certain sites are
empty and hence the future position of the trajectory.  

The proof that (1.1)-(1.3) hold for these models was given by Kipnis and
Varadhan [5] for the reversible case: if $d\ge 2$, then $D(\rho) > 0$ when
$\rho < 1$; if $d=1$, then $D(\rho)=0$ unless $p({r}) > 0$ for $|{r}| >
1$. For the non-reversible case satisfying (2.12) the result is due to
Varadhan [15].  As already mentioned, in $d = 1$, with jumps limited to
nearest neighbor sites, $p(1) = p(-1) = {1 \over 2}$, the mean square
displacement only grows like $\sqrt t$ [8].

In the asymmetric case, with $p(1) = p, 
\quad  p(-1) = 1-p$, $p > {1 \over 2}$, (1.1) 
still holds after subtraction of the drift, i.e.\  for $\hat X(t) = 
X(t) - vt$ , where the velocity $v$ is given by, $v = (2p-1)(1-\rho)$.
This was proven for $d=1$ by Kipnis [16] and  for $d \geq 3$ by Varadhan and
Yau [17] who also prove (1.3) (there is no proof for $d=2$).
Somewhat surprisingly the diffusion constant for $\hat X(t)$ 
in $d=1$ is equal to the drift $D(\rho)=(1-\rho)(2p-1)$ [16].  

The dependence of $D$ on $\rho$ is not known
even for the simple exclusion process. It is intuitively clear
that $D(\rho) \to 0$ as $\rho \to 1$. Varadhan [18]
proved that the so-called ``correlation factor'' $D(\rho)/[1 - \rho]$ is
a decreasing function of $\rho$ bounded away from zero as $\rho\to 1$.
This confirms the behavior found in numerical results for
nearest neighbor jumps [19]. He also showed that $D(\rho)$ is 
continuous in all dimensions and that for $d \geq 3$, $D(\rho)$ is 
Lipshitz, e.g. $|D(\rho)-D(\rho^\prime)| < c|\rho - \rho^\prime|$.

{\bf Remark:} We remark here for completeness that many authors (beginning
with Einstein) have studied situations where there is a special particle
with a different dynamics than the other particles of the system: in
particular, the case where an external field acts only on the special
particle (see [1]). Though we will not discuss this problem here in any
detail, we want to mention that recently, Landim, Olla and Volchan [20]
have studied a one dimensional system where the special particle jumps with
probability $p>1/2$ to the right and $1-p$ to the left -with an exclusion
rule- while all the other particles follow a symmetric dynamics with
$p=1/2$ (recall that the self diffusion constant is 0 in this system [8]).
They showed that $X_{\e}(t)$ converges in probability to a number $v(t)$
which solves a differential equation and depends on the initial macroscopic
density profile. Their result holds for a large class of initial
profiles. For instance, when the initial measure is a product Bernouilli
measure of density $\rho$, they showed that
$$
v(t)=(2p-1){1-\rho\over \rho}\sqrt{{2\over \pi}}\sqrt t.
$$

\medskip
\noindent {\bf 3.  Long Range Jumps}

We discuss now the situation where particles make jumps to a symmetric
neighborhood $U$, containing $N$ sites, with equal probability, $p({r}) =
N^{-1}$ for ${r} \in U$, $p({ r}) = 0$ otherwise. We shall be interested in
the behavior of the diffusion constant $D_N(\rho)$ and the suitably scaled
trajectory $X^N(t)$ when $N$ becomes large.  Intuitively, as $N$ increases
the tag is less likely to revisit, during a fixed number of jumps, a
previously occupied site and hence there will be less and less memory left
of previous jumps.  The only effect of the hard core exclusion will then be
to slow down the jump rates by a factor $(1 - \rho)$, the fraction of
attempted jumps which are unsuccessful due to the target site being
occupied.  Intuitively this will lead to a density independent correlation
factor, $D_N(\r)/(D_N(0)(1-\r))$, in the limit $N \to \infty$: this limit
is analogous to the van der Waals or mean field limit in equilibrium
systems when the particles interact via a long range Kac potential [21].

Since our dynamics is reversible the result of Kipnis and Varadhan [6]
applies, so that for each $N$
$$
{\buildurel {\epsilon \to 0} \under  \lim} 
{\epsilon X^N(t\epsilon^{-2})\over \sqrt {D_N(\rho)}} = W_1(t),
\eqno(3.1)
$$
and one expects that 
$$
\lim_{N\to \infty} {D_N(\rho) \over D_N(0)} = 1 - \rho.
\eqno(3.2)
$$
Actually, we show more:
\vskip1mm
\noindent{\bf Proposition.} Set $ C_N(\rho)=D_N(0)(1-\r)-D_N(\r)$.  
Then there are positive constants $c_1$ 
and $c_2$ such that 
$$
c_1 
\le F_N(\rho) =  {NC_N(\r)\over \r(1-\r)D_N(0)}\le c_2, 
\qquad\forall \r\in [0,1].\eqno(3.3)
$$

The proof of the proposition is given in the Appendix.  

In order to determine the behaviour of $D_N(\r)$, we have performed
numerical simulations in one and two dimensions with periodic boundary
conditions at different densities. The average of $[X^N(t)]^2$ over many
realizations (from 100 to 1000) was plotted against $t$ and fit to a
straight line passing through the origin. The fitted slope is then taken
for the diffusion coefficient.

The one dimensional lattice had 2000 sites.  Typically, 200 realizations
were run for 1000 time steps.  The maximum $N$ considered was 100. In two
dimensions, square and triangular lattices were used, with $200\times 200$
lattice sites.  300 realizations were run for about 500 time steps and
$N\le 90$.  Simulations with larger $N$ were also run, but the statistical
accuracy was not enough to extract any information beyond the zero order
one.  In Fig.\ 1 we plot $F_N(\rho)$ vs.\ $\rho$ for the square,
triangular, and one dimensional lattices for different values of $N$: the
set $U$ to which the particle jumps being the $N$ closest neighbors as
measured by the number of bonds required to reach the site. It appears that
$F_N(\rho)$ varies linearly in $\rho$ with a slope independent of
dimensions.

\bigskip
\noindent
{\bf Acknowledgments}

We thank C. Landim, S. Olla, M.S. Ripoll and H.T. Yau for useful discussions.  
This work was supported by NSF Grant 92-13424 4-20946. R.B. was also
supported by D.G.I.C.y T.(Spain), project PB94-0265.

\bigskip
\noindent{\bf Appendix.   Proof of the Proposition}
\vskip1mm
We will work in the moving frame of the tag particle and for simplicity
write the proof for cubic lattices. Thus, $U=\{y\not= 0\colon ||y||\le n\}$
and $N=(2n+1)^d-1$. The generator of this process is $-A$ with
$$
Af(\eta)=\sum_{y\in U}
{1\over N}(1-\eta(y))(f(\eta)-f(\tau_y\eta))
+{1\over 2} \sum_{x\in Z^d\backslash\{0\}}\  \sum_{y\in x+U}
{1\over N}(f(\eta)-f(T_{xy}\eta)),
$$
where $T_{xy}\eta=\eta^{x,y}$, $\tilde \tau_y$ 
shifts the configuration by a vector $y$,
and we have denoted $\tilde \tau_y(T_{0y})$ by $\tau_y$.
The process seen from the tag particle is reversible
with respect to $\hat\nu_{\rho}(\eta)=\nu_{\rho}(\eta|\eta(0)=1)$ 
with $\rho$ in $[0,1]$. The 
expectation with respect to $\hat\nu_{\rho}$ is denoted by $\E$.

It is well known, see [1], that 
if $S_t$ is the semigroup generated by $A$
$$\eqalign{
D_N(\rho)&=D_N(0)(1-\rho)-\int_0 ^{\infty}(S_tw,w)dt,\cr
D_N(0)&={1\over N}\sum_{||y||\le n} y_1^2\sim n^2 \qquad\text{and}\qquad
w(\eta)={1\over N}\sum_{||y||\le n, y_1>0} 
y_1(\eta (\overline{y})- \eta (y)),\cr}
$$
where $y=(y_1,\dots,y_d)$, $\overline{y}=(-y_1,y_2,\dots,y_d)$ and
because the diffusion matrix is diagonal, we chose
$w$ to be the current in the $e_1$ direction. Note that $C_N(\rho)=
-\int (S_tw,w)dt$.
\vskip1mm
We introduce the normalized variables 
$$
r_y={(\eta(y)-\r)\over \sqrt{\r(1-\r)}}\qquad
\text{such that}\qquad
\E[r_xr_y]=\d_{x,y}\qquad\forall x,y\in Z^d\backslash\{0\}.
$$
\noindent{\bf a)}  $C_N(\r)\le {\overline c}\r(1-\r){n^2\over n^d}.$

Recalling a variational formula used in [17]
$$
C_N(\r)=\sup_{f\ local}{\E[wf]^2\over \E[fAf]},
$$
we just need to show that for any local function $f$,
$$
\E[wf]^2\le {\overline c}\r(1-\r){n^2\over n^d}\E[fAf].
$$
Now,
$$
\eqalign{
\E[wf]=&{\sqrt{\r(1-\r)}\over N}
\sum_{||y||\le n, y_1>0}y_1\E[(r_{\overline{y}}-r_y)f]\cr
=& {\sqrt{\r(1-\r)}\over N}\sum_{||y||\le n, y_1>0}
y_1\E[ r_y(T_{\overline{y},y}f-f)].\cr}
$$
First, we fix $(y_2^o,\dots,y_n^o)$ and work on the line
$y=(y_1,y_2^o,\dots,y_n^o)$ for $y_1\in [-n,n]$. 
For each $y_1>1$, there are
$y_1-1$ different ways of joining $\overline{y}$ and $y$ in three steps 
$(\overline{y},ke_1+\overline{y},(k,y_2^o,\dots,y_n^o),y)$ 
with $k=1,\dots,y_1-1$ while remaining on the line
joining $y$ and $\overline{y}$. It is important to note
that $ke_1+\overline{y}$ is at $y_1$ units from $(k,y_2^o,\dots,y_n^o)$,
and thus as $y_1$ ranges from 
$2$ to $n$, a given pair of sites on the same line will be used at most once.
Now, with the notations $k_1=ke_1+\overline{y}$ and $k_2=
(k,y_2^o,\dots,y_n^o)$
$$
\eqalign{
T_{\overline{y},y}-\i&=T_{\overline{y},k_1}
T_{k_1,k_2}T_{k_2,y}T_{k_1,k_2}T_{\overline{y},k_1}-T_0\cr
&=T_1T_2T_3T_4T_5-T_0=\sum_{m=1}^5[T_0\dots T_{m-1}](T_m-T_0).\cr}
$$
Also, the product measure being invariant under exchanges
$$
\forall m\in [1,5]\qquad
\E r_y [T_0T_1\dots T_{m-1}](T_mf-f)=\E r_{y_m}(T_mf-f),
$$
where $y_m$ belongs to $\{\overline{y},k_1,k_2,y\}$. 

Now, if $y_1,y_2$ belong to a line parallel to $e_1$,
we want to `split' $T_{y_1,y_2}$ -which arose in our previous decomposition-
into $T_{y_1,z}$ and $T_{z,y_2}$ where $z$ is a common neighbor.
$y_1$ and $y_2$ 
will have of the order of $n^d$ common neighbors if $||y_1-y_2||\le n$.
However, for each neighbor, say $z$, the pair $(y_1,z)$ will be used
at most $2n$ times because in our line, $y_1$ has $2n$ 
neighbors at most. Thus, we end up with
$$
\sum_{||y||\le n, y_1>0}
y_1\E |r_y(T_{\overline{y},y}f-f)|\le {\overline {a}\over n^{d-1}}
\sum_{\scriptstyle ||x-y||\le n\atop\scriptstyle ||y||\le n}
\E |r_{\g}(T_{x,y}f-f)|
$$
where $\overline {a}$ depends just on the dimension $d$, and
$\g$ is an inocuous index different for each pair $(x,y)$.

Denoting by $T_m$ a generic exchange operator and
$\g$ an arbitrary index and using the Cauchy inequality
$$
|\E[ r_{\g}(T_mf-f)]|\le (\E[ r_{\g}^2]\E[(T_mf-f)^2])^{1/2}=
\sqrt{ \E[(T_mf-f)^2]}.
$$
gives
$$
\eqalign{
\E[wf]^2&\le [\overline {a}{\sqrt{\r(1-\r)}\over N}
\sum_{\scriptstyle ||x-y||\le n\atop\scriptstyle ||y||\le n}
{1\over n^{d-1}}
\sqrt{ \E[(T_{x,y}f-f)^2]}]^2\cr
&\le \r(1-\r)\overline {a}^2{1\over N^2}[{1\over n^{d-1}}]^2(n^d)^2
\sum_{\scriptstyle |x-y|\le n\atop\scriptstyle x,y\not= 0}
\E[(T_{x,y}f-f)^2]\cr
&\le {\overline c}{n^2\over N}\r(1-\r) \E[fAf],\cr}
$$
and the first inequality follows.

\noindent{\bf b)}  $C_N(\r)\ge {\underline c}\r(1-\r){n^2\over n^d}.$

Choose $f=\sum_{||y||\le n,y_1>0} r_y$, and write 
$$
\E[fAf]\le {1\over N}[ \sum_{0<||y||\le n} \E[(\t_yf-f)^2]+
{1\over 2}\sum_{\scriptstyle ||x-y||\le n\atop\scriptstyle x,y\not= 0}
\E[(T_{x,y}f-f)^2]].
$$
The reason to choose such an $f$ is that
$$
\E[wf]^2= {\r(1-\r)\over N^2}[(2n)^{d-1}\sum_{y_1=1}^n y_1]^2\ge 
{n^2\r(1-\r)\over 8},
$$
Now, $\t_x r_y$ is equal to $r_{x+y}$ if $y\not=-x$ and to $r_x$ if $y=-x$,
so, for any $y$, it is easy to see that
$$
\E[(\t_yf-f)^2]\le N.
$$
Thus,
$$
\sum_{0<||y||\le n} \E[(\t_yf-f)^2]\le N^2.
$$
Now, it is also easy to see that 
$$
\sum_{\scriptstyle ||x-y||\le n\atop\scriptstyle x,y\not= 0}
\E[(T_{x,y}f-f)^2]\le N^2,
$$
and thus
$$
C_N(\r)\ge {n^2\over N}\r(1-\r){1\over 16}.
$$
which completes the proof.\hfill$\qed$
\vskip1mm

\noindent{\bf References}
\vskip1mm
\frenchspacing
\item [1] H.Spohn, {\it Large Scale Dynamics of Interacting
Particles}, Springer-Verlag, 1991;
J.L.Lebowitz, H.Spohn, J.Stat.Phys.{\bf 28},539,1982.
\item [2] J.Quastel, Comm.Pure and App.Math.40,623-679(1992)
\item[3] F.Spitzer, J.Math.Mech.{\bf 18}, 973 (1968).
\item [4] D.Durr, S.Goldstein, J.L.Lebowitz, Phys.Rev.Lett., {\bf 57},
1986;  Comm.Pure and App.Math{\bf 38},573,1985.
\item [5] J.L.Lebowitz, J.Percus, and J.Sykes, Phys.Rev.{\bf
188}, 487, 1969; J.L.Lebowitz and J.Sykes, J.Stat.Phys.{\bf 6},
157--171, 1974.
\item [6] C.Kipnis, S.R.S.Varadhan, Comm.Math.Phys. {\bf 106},1-19 ,1986;
M.Z.Guo, G.Papanicolaou, Proc. Taniguchi Symp., Kyoto,1985.
\item [7] A.de Masi, P.Ferrari, S.Goldstein, D.Wick, J.Stat.Phys.
55,787-855,1985.
\item [8] R.Arratia, Ann.Prob,11,362-373,1983.
\item [9] H.J.Bussemaker, J.Dufty, M.Ernst, J.Stat.Phys, 78:1521(1995);
R.Esposito, R.Marra, H.T.Yau, Rev. Math. Phys. 1995.
\item [10] R.Fernandez, A.Sokal, A.van Enter, J.Stat.Phys. {\bf 72},
879--1167, 1993.
\item [11] F.Spitzer, Adv.Math.,5,246-290,1970.
\item [12] E.Saada, Ann.Inst.H.Poincar\'e, Prob-Stat. 26(1),5-17,1990.
\item [13] P.Siri, Ph.D. thesis, Torino, 1996.
\item [14] H.van Beijeren, J.Stat.Phys.60,845,1990.
\item [15] S.R.S.Varadhan, Ann.Inst.H.Poincar\'e,Prob-Stat,31,273-285,(1995).
\item [16] C.Kipnis, Ann.Prob,Vol 14,1986,pp397-408.
\item [17] S.R.S.Varadhan, H.T.Yau, private communication, 1996.
\item [18] S.R.S. Varadhan, preprint 1993.
\item [19] K.Kerr, K.Binder, in {\it Application of the Monte Carlo method
in Statistical Physics} (Berlin,Springer 1984). 
\item [20] C.Landim, S.Olla, S.B.Volchan, preprint 96.
\item [21] G.Giacomin and J.L.Lebowitz, Phys.Rev.Lett.{\bf
76},1094-1097,1996. 
\vfill \eject
{\bf Fig.1} Simulation results of the  quantity $F_N(\rho)$ 
versus the density $\rho$.  Results of simulations in one dimensional 
lattices are represented by circles, in a square lattice by squares
and in the triangular lattices by triangles. The number of neighbours
is shown in the legend. The slope of $F_N(\rho)$ as a function
of $\rho$ appears to be  independent of the dimension or the type of
lattice. 
The intercept, on the contrary, appears to depend on the dimension
but not on the lattice structure. 
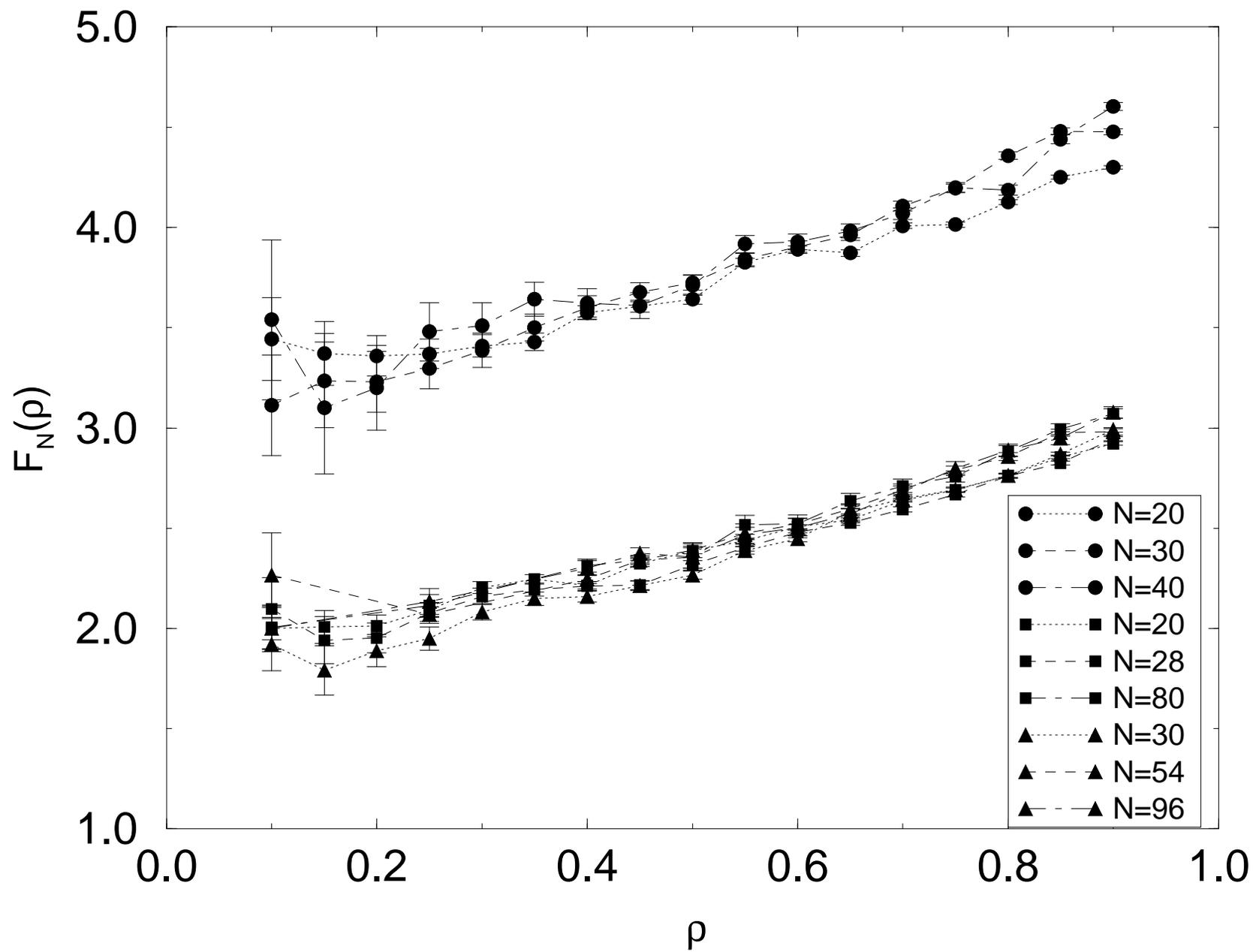
\bye